# A comprehensive first principles calculations on $(Ba_{0.82}K_{0.18})(Bi_{0.53}Pb_{0.47})O_3$ single-cubic-perovskite superconductor

M. H. K. Rubel[1*], Sujon Kumar Mitro[2], Khandaker Monower Hossain[1], M. Mijanur Rahaman[1,**], M. Khalid Hossain[3], Jaker Hossain[4], B. K. Mondal[4], Istiak Ahmed[5], A. K. M. A. Islam[5,6], A. El-Denglawey[7]

[1]Dept. of Materials Science and Engineering, University of Rajshahi, Rajshahi 6205, Bangladesh
[2]Bangamata Sheikh Fojilatunnesa Mujib Science and Technology University, Jamalpur, Bangladesh
[3]Bangladesh Atomic Energy Commission, Dhaka 1349, Bangladesh
[4]Dept. of Electrical and Electronic Engineering, University of Rajshahi, Rajshahi 6205, Bangladesh
[5]Dept. of Physics, University of Rajshahi, Rajshahi 6205, Bangladesh
[6]International Islamic University Chittagong, Kumira, Chittagong 4318, Bangladesh
[7]Dept. of Physics, Taif University, P.O. box 11099 Taif 21944, Saudi Arabia

Correspondence: *mhk_mse@ru.ac.bd; **mijan_mse@ru.ac.bd

**Abstract**

In this present study, the pseudopotential plane-wave (PP-PW) pathway in the scheme of density functional theory (DFT) is utilized to investigate the various physical properties on $(Ba_{0.82}K_{0.18})(Bi_{0.53}Pb_{0.47})O_3$ (BKBPO) single perovskite superconductor. We have analyzed elastic constants and moduli at zero and elevated pressures (up to 25 GPa) as well. The calculated values of Poisson's ratio, Pugh's indicator, and Cauchy pressure of the studied superconductor are found to be brittle in nature at ambient conditions. The calculated machinability index and hardness values of BKBPO perovskite indicate its superior industrial applications to similar ones. We also have investigated the anisotropic nature incorporating both the theoretical indices and graphical representations in 2D and 3D dimensions, which reveals a high level of anisotropy. The flatness of the energy bands near $E_F$ is a sign of Van-H$_f$ singularity that might increase the electron pairing and origination of high-$T_C$ superconductivity. The computed band structure exhibits its metallic characteristics is confirmed by band overlapping. A band of DOS is formed for the strong hybridization of the constituent elements. The orbital electrons of O-$2p$ contribute most dominantly at $E_F$ in contrast to all orbital electrons. The orbital electrons at the $E_F$ are higher from both the partial density of states and charge density mapping investigation. The coexistence of the electron and hole-like Fermi sheets exhibits the multi-band nature of BKBPO. On the other hand, Fermi surfaces with flat faces promote transport features and Fermi surface nesting as well. The calculated value of the electron-phonon coupling constant ($\lambda = 1.46$) is slightly lower than the isostructural superconductor, which indicates that the studied BKBPO can be treated as a strongly coupled superconductor similar to the reported isostructural perovskite superconductors. Furthermore, the thermodynamic properties have been evaluated and analyzed at elevated temperature and pressure by using harmonic Debye approximation (QHDA).

## 1 Introduction

Perovskite oxide (ABO$_3$) is one of the most useful classes of materials for its unique physical properties, such as superconductivity (high-$T_c$) [1], ferroelectricity [2], piezoelectricity [3], colossal magnetoresistivity [4], and ion conductivity [5–8]; as well as their suitability in numerous technological applications, which include modulators and devices [9], detectors and sensors of infrared energy [10], memory devices [11], high-frequency microwave capacitors [12], and so on [13]. In the midst of them, perovskite bismuth oxides have achieved unprecedented attention by researchers in recent years owing to their diverse interesting physical attributes, namely, superconductivity, ferroelectricity, and ferromagnetism by partially substituting either A- and/or B-site elements in their crystal motifs [14,15]. Among these properties, superconductivity is phenomenal and at the beginning Sleight et al. explored superconducting behavior in $BaBiO_3$ by partially substituting Bi with Pb that yielded transition temperature ($T_c$) of 13 K [16]. After that, Cava et al. [17,18] and Mattheiss et al. [19] reported a new superconducting perovskite bismuthate $Ba_{1-x}K_xBiO_3$ with a higher $T_c$ of ~30 K. The invention of new superconducting material with high $T_c$ is regarded as a challenge for the researchers. The above research works further inspired to discovers several Bi-based superconductors $(Na_{0.25}K_{0.45})Ba_3Bi_4O_{12}$ [20], $(K_{1.00})(Ba_{1.00})_3(Bi_{0.89}Na_{0.11})_4O_{12}$ [21], $(Ba_{0.62}K_{0.38})(Bi_{0.92}Mg_{0.08})O_3$ [22], $(Ba_{0.54}K_{0.46})_4Bi_4O_{12}$ [23], and $(Ba_{0.82}K_{0.18})(Bi_{0.53}Pb_{0.47})O_3$ [24]. Among these, superconducting bismuth oxide $(Ba_{0.82}K_{0.18})(Bi_{0.53}Pb_{0.47})O_3$ (BKBPO) was successfully synthesized by Rubel et.al by using an environment-friendly low-temperature hydrothermal reaction method, which exhibited a $T_c$ of ~22.8 K [24] in the family of Bi/Pb perovskite. The refinement results based on SXRD data shows the simple perovskite-type structure (S.G.: $pm\bar{3}m$, No. 221) of the compound. It revealed a diamagnetic sign with 64% volume fraction of superconductivity and zero resistivity occurred at ~4 K. Although, only the electronic band structure of the superconductor was predicted with experimental observations [24], but there are still huge scopes to explore the delicate learnings about the material system. Altogether, the theoretical study on its physical properties is yet available in the existing literature. Moreover, four of the previously discovered Bi-based perovskite superconductors were thoroughly investigated by the first-principles method and demonstrated some interesting findings [25–28]. Thus, it is really logical to show circuity for the investigations of physical characteristics of the BKBPO superconductor. Recently, first-principle calculations based on the density functional theory (DFT) have become a crucial technique to explore the physical properties of solid materials [29–37]. Hence, the prime objective of our study is to grasp the structural and mechanical properties, elastic anisotropy, electronic behaviors (band diagram, density of states, Fermi surface, and electronic charge density mappings), thermodynamic properties (bulk modulus, Debye temperature, specific heat, volume thermal expansion coefficient, and entropy) at zero and elevated temperature, and electron-phonon coupling constant of BKBPO superconductor, which could open oncoming science base investigations and potent applications of this material together with other strongly related systems. In the present scheme, we have employed Cambridge Serial Total Energy Package (CASTEP) in the context of DFT to analysis the ingoing physical attributes and Gibbs program to explore the thermodynamic properties of BKBPO for the first time; and the computed results are compared with the available data of other similar superconducting perovskites.

## 2 Computational Methodology

The current research uses the CASTEP package [38] employing the pseudo-potential plane-wave (PP-PW) full energy estimation of the DFT method [39]. The crystal structure of $(Ba_{0.82}K_{0.18})(Bi_{0.53}Pb_{0.47})O_3$ superconductor is constructed using the experimentally reported crystallographic data [24]. This compound has a cubic perovskite lattice of $Pm\bar{3}m$ with, $a$ = 4.28877 (1) Å. In the unit cell, both Ba and K atoms are randomly distributed at the corners with the Wyckoff position of 1$b$ (0.5, 0.5, 0.5); Bi and Pb atoms are randomly located at the corners having the Wyckoff position, 1*a* (0, 0, 0); and O atoms hold the edge centers position at 3*d* (0.5, 0, 0). The electronic exchange-correlation is arranged in the framework of Generalized Gradient Approximation (GGA) assembled with Perdew-Burke-Ernzerhof (PBE) [40]. The Vanderbilt-type ultrasoft pseudopotential [41] has been accounted for the electron-ion interactions, whereas, Broyden-Fletcher-Goldfarb-Shanno (BFGS) technique [42] is used for the optimization of BKBPO structure. The plane wave cut-off energy was fixed to 500 eV together with the *k*-point mesh of 12×12×12 using the outline of the Monkhorst-Pack scheme [43]. But, in the calculation of Fermi surface and charge density mapping, a very high *k*-point mesh of 22×22×22 is imposed to achieve better precision. In addition, the sampling integration is settled to the ultrafine state throughout the first Brillouin zone in the crystal lattice, where the atomic positions, as well as lattice parameters, retain quite relax [44,45] during the optimization process. Converging parameters for the geometrical optimization bears the magnitudes as follows: total energy within $5\times10^{-6}$ eV/atom; maximum ionic force within 0.01 eV/Å, maximum ionic displacement within $5.0\times10^{-4}$ Å, and maximum stress of 0.02 GPa. In addition, the virtual crystal approximation (VCA) is also exploited to optimize the unit structure, which is an alternative to the super cell scheme convenient for estimating the physical properties of disordered materials. Employing such method, K and Pb atoms are arbitrary mixed with Ba and Bi atoms, respectively. The VCA scheme was effectively utilized to numerous disordered systems, such as silicates, perovskites, ferroelectric ceramics, superconductors, and MAX phases [28,33,46,47] already. The mechanical and electronic features are also calculated by considering the aforementioned parameters. The elastic moduli are determined by the 'stress-strain' method [48] incorporated in the CASTEP package. Moreover, to calculate the thermodynamic properties of BKBPO, the quasi-harmonic Debye model (with the vibrational term) was implemented in the Gibbs program [49]. Under this scheme, the bulk modulus, Debye temperature, specific heat, volume thermal expansion coefficient, and entropy are calculated at 0-1000 K temperature and 0-50 GPa pressure using the DFT calculated E-V data (at zero temperature and zero pressure) utilizing the Birch–Murnaghan equation of state [50].

## 3    Results and discussion

### *3.1    Structural and mechanical properties*

A novel Bi-based simple perovskite superconductor BKBPO with space group $Pm\bar{3}m$ (#221) having cubic structure [24] with atomic coordinates A (0.5, 0.5, 0.5), B (0.5, 0, 0) has been chosen for the study. In the drawn structure (by VESTA), the A-site is partially covered by K and Ba atoms, whereas B-site is partly occupied by Pb and Bi, and O-site is completely filled with O-atoms. The geometry optimization was carried out by CASTEP code including the generalized gradient approximation (GGA). The simple cubic structure of BKBPO is optimized at zero pressure under ambient conditions and presented in **Figure 1**. After successful optimization, the consistent lattice parameter of BKBPO is taken and listed in **Table 1**, which fairly agrees with the experimental outcomes [22,24,27]. Moreover, **Table 1** displays the data on calculated and experimental lattice parameter (*a*) of isostructural simple cubic perovskite [22,24,27] for comparisons. However, the influence of spin-orbit coupling (SOC) in our studied superconductor is not considered in DFT Hamiltonian [14,15] owing to the diamagnetic nature and complex composition feature of the material. Although several researchers have included SOC effect in DFT-method in Bi-based superconductors which results in a very minor effect on the lattice parameter, total

DOS as well as electronic states near $E_F$. Therefore, regarding the aforementioned issues, the calculation in this study is simply carried out without SOC.

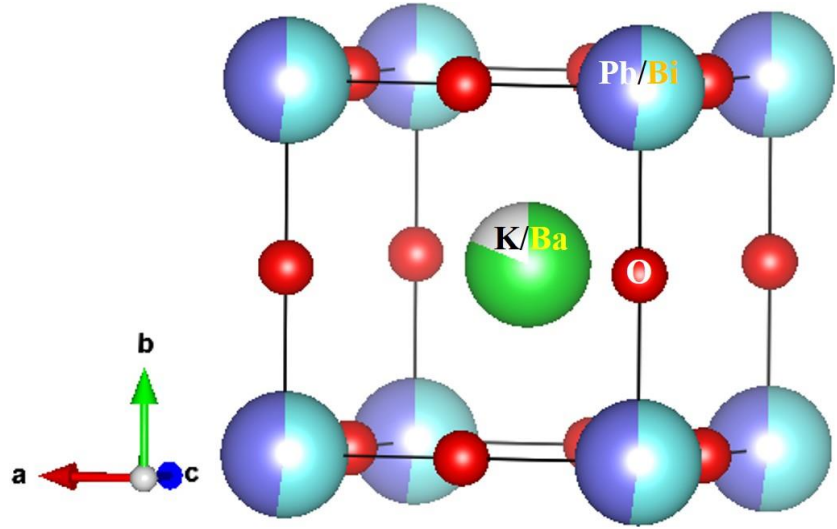

**Figure 1.** The optimized crystal structure of the BKBPO superconductor.

**Table 1.** Calculated lattice ($a$ in Å) and elastic constants ($C_{ij}$ in GPa), bulk modulus ($B$ in GPa), shear modulus ($G$ in GPa), Young's modulus ($E$ in GPa), Pugh's indicator ($G/B$ in GPa), machinability index($\mu_M$), Poisson's ratio ($v$), and Vickers hardness ($H_v$ in GPa) of BKBPO at $P = 0$ GPa.

| Compound | $a$ | $C_{11}$ | $C_{12}$ | $C_{44}$ | $B$ | $G$ | $E$ | $G/B$ | $\mu_M$ | $v$ | $H_V$ | Ref. |
|---|---|---|---|---|---|---|---|---|---|---|---|---|
| BKBPO | 4.16086 | 168 | 100 | 72 | 123 | 53 | 139 | 0.43 | 1.72 | 0.311 | 6.68 | This |
|  | 4.28877 | - | - | - | - | - | - | - | - | - | - | [24] |
| BKBMO | 4.20300 | 192 | 32 | 58 | 85 | 65 | 155 | 0.76 | 1.46* | 0.190 | - | [27] |
|  | 4.27900 | - | - | - | - | - | - | - | - | - | - | [22] |

BKBPO: $(Ba_{0.82}K_{0.18})(Bi_{0.53}Pb_{0.47})O_3$, BKBMO: $(Ba_{0.62}K_{0.38})(Bi_{0.92}Mg_{0.08})O_3$. *Calculated by us.

The elastic constants of a solid substance reveal its reply to external stress as well as certain boundary factors are associated with the workable applications of a product [51]. Therefore, these moduli are essential in the view of materials science and engineering for application purposes. By inquiring about the elastic-mechanical behaviors of a solid material we can predict the tolerance ability of applied stress for different given environments [51]. Moreover, the deep sense of the elastic constants of material is very crucial for many practicing applications adhered to the mechanical behavior of the solid such as thermo-elastic stress, load deflection, internal strain, fracture toughness, and sound velocities. In order to accomplish the elastic and mechanical characteristics fairly for cubic crystal merely three self-sufficient independent elastic constants ($C_{11}$, $C_{12}$, and $C_{44}$) are necessary [52]. Besides, both stress and strain have three tensile and three shear components, producing six components totally. The polycrystalline mechanical properties such as Bulk modulus ($B$), Shear modulus ($G$), Young's modulus ($E$), and Poisson's ratio ($v$) can be calculated from these independent elastic constants. The researcher, Hill et. al. [53] proved that the Voigt and Reuss equations represent upper and lower limits of the real polycrystalline constant. The authors showed that the polycrystalline moduli are the arithmetic values of the moduli in the Voigt ($B_V$, $G_V$) and Reuss ($B_R$, $G_R$) approximation, and are given by $B_H \equiv B = ½(B_R + B_V)$ and $G_H \equiv G = ½(G_R + G_V)$, where $B$ and $G$ represent the Bulk and Shear moduli, respectively. The Young's modulus, $E$, and Poisson's ratio, $v$ are then calculated from these data using the following formulas (**Eq. 1**) [54,55]:

$$E = \frac{9BG}{3B+G} \text{ and } v = \frac{(3B-2G)}{[2(3B+G)]} \tag{1}$$

The Young's modulus, $E$ of above expression is inversely connected with critical thermal shock coefficient, $R = \frac{\sigma(1-v)}{\alpha E}$ where, σ is the flexure strength and α is the coefficient of thermal expansion. The application of material as TBC is strongly dependent on the value of R. It is seen from **Table 1** that the $E$ of BKBPO is fairly lower than similar superconductors [25,26]. This smaller value of $E$ will yield larger value of $R$ makes this material a promising candidate for TBC application than the previous report [27]. Furthermore, the values of elastic constants $C_{ij}$ have been calculated by using the "stress-strain method" at zero pressure of the superconducting compound BKBPO and are listed in **Table 1**. The Born stability criteria for the mechanical permanency of cubic crystal are as follows (**Eq. 2**) [56]:

$$C_{11} > 0, C_{12} > 0, (C_{11} - C_{12}) > 0, \text{ and } C_{44} > 0 \tag{2}$$

Our calculated elastic constants of BKBPO completely obey the above conditions under 0 GPa indicating that BKBPO is a mechanically stable alloy. The calculated values of the bulk modulus ($B$), shear modulus ($G$), Young's modulus ($E$), Poisson's ratio ($v$), and Pugh's indicator ($G/B$) of BKBPO at zero pressure are given in **Table 1**. For practical applications of a solid, it is necessary to classify it as either brittle or ductile by applying some well-known relations. Among these relations, Pugh and Poisson's ratios are efficacious parameters and one can easily predict either the brittle and/or ductile behavior of solids following the marginal magnitudes. Moreover, the hardness and brittleness of a material depend on the ratio between the shear modulus to bulk modulus ($G/B$). Based on the $G/B$ ratio [57], the studied superconductor BKBPO shows ductility as the separation value of $G/B$ is less than 0.57. Frantsevich *et al.* [58] used $v < 0.26$ value for brittle materials and ductile solids it is ≥0.26. Therefore, the Bi-based perovskite superconductor is found to be ductile in accordance with the above-mentioned indicator, whereas $(Ba_{0.62}K_{0.38})(Bi_{0.92}Mg_{0.08})O_3$ is brittle (**Table 1**). Importantly, the value of $v$ is small ($v = 0.1$) for covalent materials but for ionic materials, $v = 0.25$ or more which suggests that our system can be treated as an ionic alloy. On the other hand, the value of $v$ is between 0.25 and 0.5 for the central force solid. So, the studied superconductor can be considered as central force solid material ($0.25 < v < 0.50$) [59]. Besides, the Cauchy pressure is defined as ($C_{12} - C_{44}$) [60], and if it is positive, the material is expected to be ductile; otherwise brittle. Therefore, all the studied parameters conclude that our BKBPO superconductor is ductile in nature.

The machinability index, $\mu_M$ denotes the favorable economic scale of machine utilization, plasticity, temperature and plastic strain, the lubricating nature of a material, and cutting forces is written as **Eq. 3** [61]:

$$\mu_M = B/C_{44} \tag{3}$$

A solid material that has a large value of $\mu_M$ bears excellent lubricating properties, lower friction value, and higher plastic strain value [52], and our studied BKBPO superconductor has a fairly greater value than $(Ba_{0.62}K_{0.38})(Bi_{0.92}Mg_{0.08})O_3$ is a mark of a good level of machinability. The better level of machinability and ductility make BKBPO suitable for easy mechanical manipulation compared to $(Ba_{0.62}K_{0.38})(Bi_{0.92}Mg_{0.08})O_3$ simple cubic superconductor. The following (**Eq. 4**) empirical formula [29,62] correlates the Vicker's hardness with the elastic moduli to address the hardness of materials and the value of $H_V$ is 6.68 GPa (**Table 1**), which is larger than the hardness values of superconducting $(Ba_{0.54}K_{0.46})_4Bi_4O_{12}$ and some half-Heusler (ScPdBi, ScNiBi,

LuPtBi, and LuPdBi) compounds [28,51,63,64]. Therefore, the studied superconducting BKBPO is fairly good for conventional applications.

$$H_V = \frac{(1-2\nu)E}{6(1+\nu)} \qquad (4)$$

It is hard to overstate the role of pressure on the elastic constants and moduli in understanding, exploring, and modifying the solid materials for various practical applications. We have applied hydrostatic pressures on the studied BKBPO superconductor, which is ranging from 0 to 25 GPa. **Figures 2(a)** and **2(b)** demonstrate the variation of elastic constants ($C_{11}$, $C_{12}$, and $C_{44}$), and elastic moduli ($B$, $G$, and $E$), respectively, under hydrostatic pressure. It is interesting to note that all the elastic constants are increased with the increasing pressure and they are positive as well (**Figure 2(a)**). To get more knowledge about the effect of pressure on the BKBPO compound, the elastic moduli under pressure has also been computed. We see that the elastic moduli increase monotonically with pressure (**Figure 2(b)**), but the values of $C_{11}$, and $B$, vary significantly under pressure. As $C_{11}$ denotes elasticity in the *x*-direction and thus, a volume change is strongly connected to pressure [65], it yields a large change in $C_{11}$. In contrast, $C_{44}$ and $G$ increase slowly.

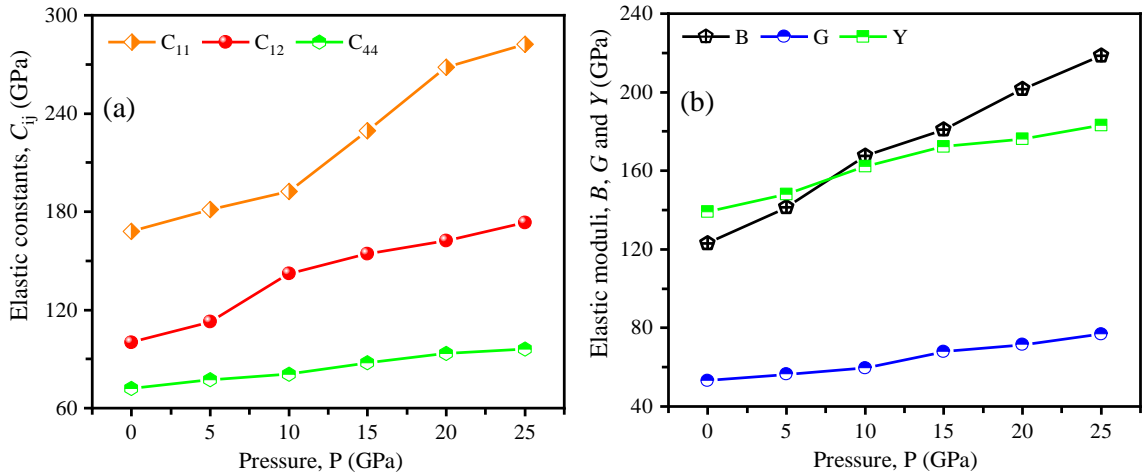

**Figure 2.** Pressure-dependent elastic (a) constants and (b) moduli of BKBPO superconductor.

## 3.2 Anisotropic nature of elastic moduli and sound velocities

The directionally dependent properties of a system can be described from the knowledge of various anisotropy indices. Anisotropy of solid material is caused by asymmetry and specific orientation of the molecules themselves in a material. Most importantly, various applications of a system under varying external stress in engineering purposes, as well as crystal physics and mechanical durability can be developed significantly by studying the elastic anisotropy [52]. The universal anisotropic index, $A^U$ proposed by Shiva Kumar and Ranganathan et. al. can be expressed as follows (**Eq. 5**) [66]:

$$A^U = 5\frac{G_V}{G_R} + \frac{B_V}{B_R} - 6 \geq 0 \qquad (5)$$

The calculated $A^U$ value for the BKBPO superconductor indicates the anisotropic nature, as $A^U = 0$ is used for the isotropic nature of a substance. Additionally, the concept of percent anisotropy in shear and bulk modulus ($A^G$ and $A_B$) proposed by Chung and Buessem [67] can be written as follows (**Eqs. 6** and **7**):

$$A^G = \frac{G_V - G_R}{2G_H} \tag{6}$$

$$A_B = \frac{B_V - B_R}{B_V + B_R} \tag{7}$$

Here, $A_B = A^G = 1$ gives the highest elastic anisotropy, whereas, $A_B = A^G = 0$ represents the elastic isotropy of a material [67]. The observed value of $A^G$ indicates the dominant shear anisotropy, whereas, the bulk modulus is isotropic (as for cubic crystal $B_H = B_R = B_V$) for the compound under study. To get more specific results in understanding the elastic anisotropy, the shear anisotropy in different directions at various crystallographic planes has also been studied and discussed. The shear anisotropic factors $A_i$ ($i$ = 1 to 3 i.e., $A_1$, $A_2$, and $A_3$) are defined as follows (**Eqs. 8-10**) [67,68] and enlisted in **Table 2**.

for {100} shear plane,
$$A_1 = \frac{4C_{44}}{C_{11} + C_{33} - 2C_{13}} \tag{8}$$

for the {010} shear plane,
$$A_2 = \frac{4C_{55}}{C_{22} + C_{33} - 2C_{23}} \tag{9}$$

for the {001} shear plan,
$$A_3 = \frac{4C_{66}}{C_{11} + C_{22} - 2C_{12}} \tag{10}$$

**Table 2.** Calculated Shear anisotropic factors $A_i$ ($i$ = 1-3), Zener's anisotropy index ($A$), anisotropy in shear ($A^G$), anisotropy in bulk modulus ($A_B$), universal anisotropy index ($A^U$), and equivalent Zener anisotropy measure ($A^{eq}$) of BKBPO superconductor.

| Compound | $A_1$ | $A_2$ | $A_3$ | $A$ | $A^G$ | $A_B$ | $A^U$ | $A^{eq}$ |
|---|---|---|---|---|---|---|---|---|
| BKBPO | 2.12 | 2.12 | 2.12 | 2.12 | 0.066 | 0 | 0.708 | 1.59 |

The level of anisotropy rises owing to the deviation from unit value, whereas, for an isotropic crystal, it is, $A_1 = A_2 = A_3 = 1$. The values of $A_i$ ($i$ = 1 to 3) show elastic anisotropy for $(Ba_{0.82}K_{0.18})(Bi_{0.53}Pb_{0.47})O_3$ superconductor, which obey our previous discussion and the relation $A_1 = A_2 = A_3$ reflects its cubic symmetry. Furthermore, Zener anisotropy index ($A$), and equivalent Zener anisotropy ($A^{eq}$) are also analyzed to get the precise anisotropy measurements by the following well-known relations (**Eqs. 11-12**) [55,69]:

$$A = \frac{2C_{44}}{C_{11} - C_{12}} \tag{11}$$

$$A^{eq} = \left(1 + \frac{5}{12}A^U\right) + \sqrt{\left(1 + \frac{5}{12}A^U\right)^2 - 1} \tag{12}$$

All the calculated values are displayed in **Table 2** and the elastic indices confirm the anisotropic nature of the studied superconductor.

The two dimensional (2D) and three dimensional (3D) direction dependency graphical representation of Young modulus ($E$), Shear modulus ($G$), and Poisson ratio ($v$) have been shown in **Figures 3(a)**, **3(b)**, and **3(c)**, respectively to understand the anisotropy more appropriately. The deviation from spherical shape indicates its degree of anisotropy at varying directions, whereas, a perfectly isotropic crystal holds the spherical shape. It is clear from **Figures 3(a-c)** that anisotropic character can be found for BKBPO superconductor and these results follow the earlier discussion using different anisotropy indices.

The observed 2D and 3D graphical plots conclude that anisotropy increases following the order $E < G < v$ (**Figure 3**). On the other hand, the maximum and minimum worth of these three entities at different directions have been given in **Table 3** for the BKBPO superconductor to compare the anisotropy in elastic moduli.

The calculated values of $E$, $G$, and $v$ in three primal directions are listed in **Table 4**, where both theoretical and graphical approaches confirm the anisotropy in elastic moduli for the BKBPO superconductor.

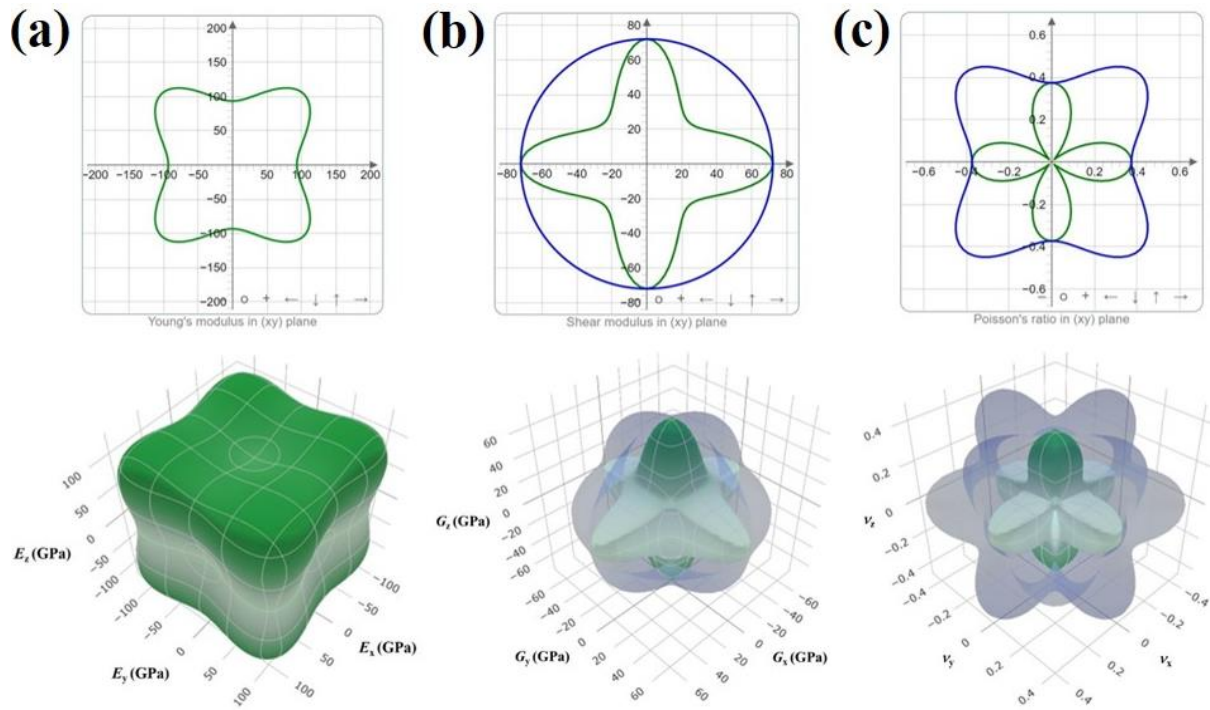

**Figure 3.** Anisotropy in (a) Young's modulus, $E$ (GPa), (b) Shear modulus, $G$ (GPa), and (c) Poisson's ratio, $v$ for BKBPO superconductor. The 2D plots are given in the upper row, whereas, 3D plots are in the lower row.

**Table 3.** The maximum and minimum limits of Young's Modulus, $E$ (in GPa), Shear Modulus, $G$ (in GPa), and Poisson's ratio, $v$ for BKBPO superconductor.

| Compound | Young's Modulus | | Shear Modulus | | Poisson's ratio | |
|---|---|---|---|---|---|---|
| | $E_{max}$ | $E_{min}$ | $G_{max}$ | $G_{min}$ | $v_{max}$ | $v_{min}$ |
| BKBPO | 130.65 | 93.37 | 72.0 | 34.0 | 0.5151 | 0.0169 |

**Table 4.** The calculated values of Young's Modulus ($E$), Shear Modulus ($G$), and Poisson's ratio ($v$) in three principal directions for cubic perovskite BKBPO superconductor.

| Compound | $E$ (in GPa) | | | $G$ (in GPa) | | | $v$ | | |
|---|---|---|---|---|---|---|---|---|---|
| | $E_x$ | $E_y$ | $E_z$ | $G_x$ | $G_y$ | $G_z$ | $v_x$ | $v_y$ | $v_z$ |
| BKBPO | 105.36 | 104.87 | 100.96 | 68.31 | 68.04 | 62.23 | 0.364 | 0.359 | 0.351 |

Another important parameter for solid is connected to acoustic wave velocities and these velocities are generally anisotropic in nature. To measure the sound velocities in different travelling directions of the studied compound, we have applied Brugger's suggested methods [70] of pure longitudinal and transverse modes for the propagation of elastic waves. For doing this, the sound velocities must be treated in [001], [110], and [111] directions, as the sound propagating modes are in the quasi-transverse or quasi-longitudinal. In the case of a cubic crystal, the acoustic wave velocities can be expressed as **Eq. 13** [71]:

$$[100]v_l = \sqrt{\frac{C_{11}}{\rho}}, [010]v_{t1} = [001]v_{t2} = \sqrt{\frac{C_{44}}{\rho}}, [110]v_l = \sqrt{\frac{(C_{11}+C_{12}+2C_{44})}{2\rho}},$$

$$[1\bar{1}0]v_{t1} = \sqrt{\frac{(C_{11}-C_{12})}{\rho}}, [001]v_{t2} = \sqrt{\frac{C_{44}}{\rho}}, [111]v_l = \sqrt{\frac{(C_{11}+2C_{12}+4C_{44})}{3\rho}}, \quad (13)$$

$$[11\bar{2}]v_{t1} = v_{t2} = \sqrt{\frac{(C_{11}-C_{12}+C_{44})}{3\rho}}$$

Where, $v_{t1}$ is the first transverse mode and $v_{t2}$ denotes the second transverse mode. The calculated acoustic velocities along different directions of BKBPO perovskite are given in **Table 5**. It is worth mentioning that, these unlike values of velocities in different directions indicate lattice dynamical anisotropy in $(Ba_{0.82}K_{0.18})(Bi_{0.53}Pb_{0.47})O_3$ superconductor.

**Table 5**. Anisotropic sound velocities (ms$^{-1}$) of BKBPO perovskite superconductor along with different crystallographic directions.

| Directions | Directional velocity modes | Magnitudes of sound velocities |
| --- | --- | --- |
| [111] | $[111]v_l$ | 4281.25 |
|  | $[11\bar{2}]v_{t1,2}$ | 1977.80 |
| [110] | $[110]v_l$ | 4155.40 |
|  | $[1\bar{1}0]v_{t1}$ | 2387.45 |
|  | $[001]v_{t2}$ | 2456.66 |
| [100] | $[100]v_l$ | 3752.62 |
|  | $[010]v_{t1}$ | 2456.66 |
|  | $[001]v_{t2}$ | 2456.66 |

## 3.3 Electronic band structure and density of states

The band structure is the spectrum of the energy eigenvalues of a periodic system that contains information about both the bonding interactions within molecules (intra-molecular) and the intermolecular interactions. However, the material properties can be understood if one can identify the character of dominant bands at the Fermi level, their energies, etc. The calculation of the electronic band structure helps to understand the shape of the Fermi surface. **Figure 4(a)** depicts the energy bands of BKBPO superconductor at zero pressure along (Γ-X-M-Γ-R-X) high symmetry directions of isostructural simple perovskite BKBMO superconductor [27] of the Brillouin zone in the energy range from -20 to +20 eV to compare. For the compound, Fermi level ($E_F$) is set at 0 eV as conventional, which is presented by the dotted line. The valence and conduction bands are clearly seen to overlap, indicating the metallic behavior of the BKBPO superconductor. The calculated band diagram in this study is very analogues to the reported BKBO perovskite superconductors [25–28]. The calculated band structure also reveals a van Hove Singularity (vHS- flatness of band) at M point adjacent to $E_F$ level which is a signature of the emergence of superconducting nature [24,27].

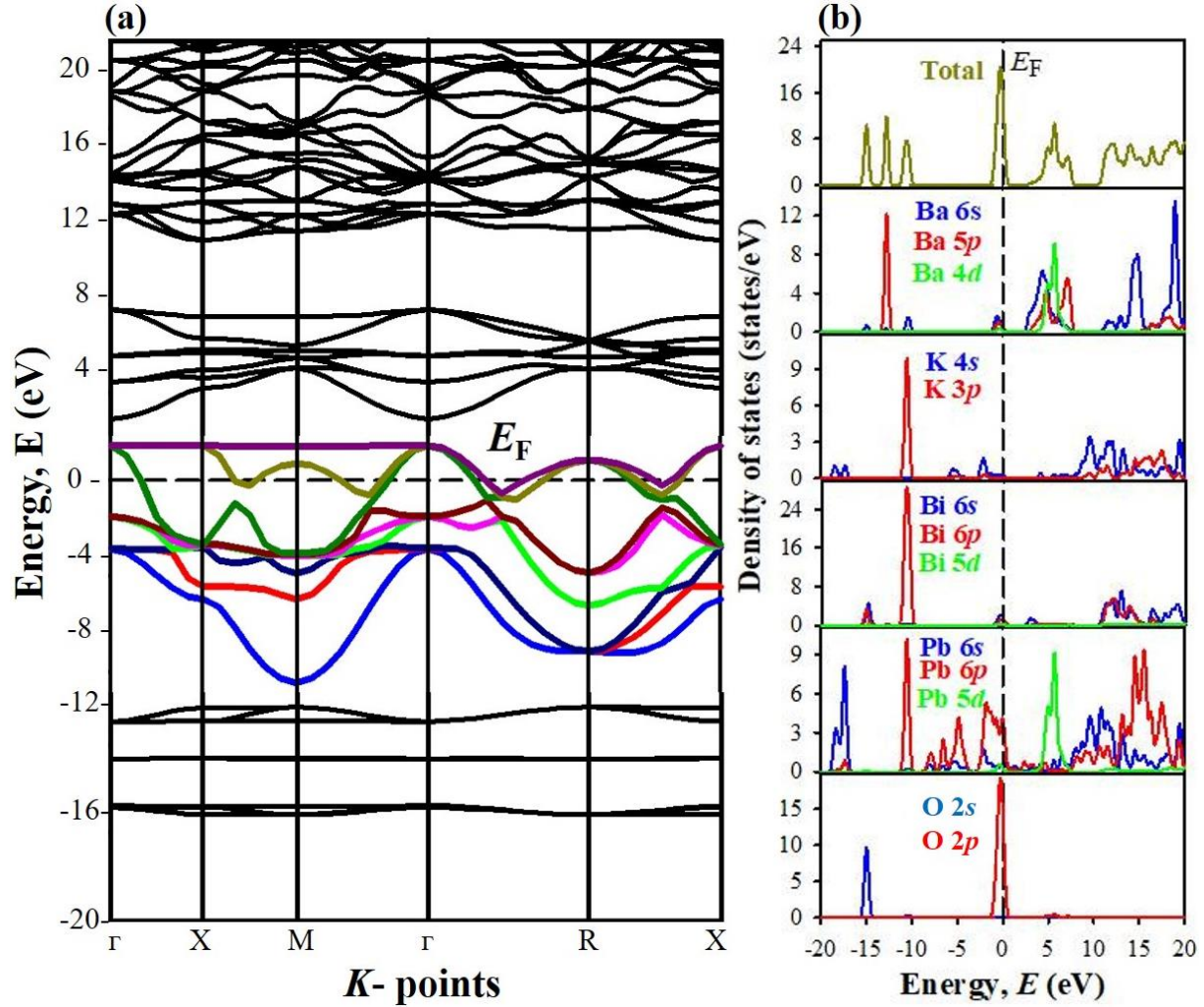

**Figure 4.** The electronic band structure (a) and DOS (b) of the BKBPO superconductor.

Besides, the density of states shows a qualitative measurement for understanding the electronic structure of any material system is defined as the number of states per interval of energy per unit volume. From **Figure 4(b)**, it is observed that peaks in the valence band within the energy range -13.0 eV to -4.0 eV of BKBPO are formed almost entirely by Ba/K-6$s$,5$p$,4$d$/4$s$,3$p$, Bi/Pb-6$s$,6$p$, and O-2$p$ states. Finally, the energy bands are mainly contributed at the $E_F$ from O-2$p$ and Pb-6$p$ with moderate contributions of Pb/Bi-6$s$, and Ba-6$s$ states. Whereas, small contributions of K-4$s$,3$p$, Ba-4$d$, and Bi/Pb-5$d$ are also observed. The PDOS curve mainly reveals a robust hybridization between p orbital electrons of Pb and O atoms together with $s$ orbital electrons of Ba/K and Bi atoms at $E_F$. The significant DOS values are seen in the energy band diagram at $E_F$ with the p-type of carriers in the material. Notably, in these DOS figures, the lowest valance bands originate from -20 to -13 eV for the hybridization of O-2$s$ along with Bi/Pb-6$s$, K-4$s$, and Ba-6$s$ orbitals electrons which is a mark of ionic bonding between Bi/Pb – O and Ba/K – O. Moreover, -12 eV to $E_F$ one sharp peak is created for the strong hybridization of $p$ orbitals of each atom, trace the valence bond among Ba/K, Bi/Pb, and O atoms. The few other relatively broadened peaks for orbital electrons of an individual atom are found owing to the topmost valance electrons.

This scenario of the DOS band structure is very identical to the isostructural BKBMO perovskite superconductor [27] of an earlier study.

### 3.4 Fermi surface and charge density mapping

To visualize the electronic behavior and carry out the evidence of band crossing the $E_F$ of a material, the Fermi surface topology is an essential investigation. From **Figure 5**, both electron and hole-like sheets are clearly seen due to the bands crossing of $E_F$ from valence and conduction states. This result is a symbol of the possible multi-band behavior of the BKBPO superconductor. In **Figure 5**, at the R point of the topology, three electrons like layered separated curve sheets (two are close and small whereas, the other one is far and relatively bigger) are seen. The R-Γ line meets a solid ball-shaped hole-like sheet from the R point by crossing through three electron sheets. Besides, the R-X line touched the edges of three electron sheets only and X-point stands in free space. Moreover, a total of 24 electrons-like and only one hole-like sheet are observed in the unit cell of the BKBPO superconductor. On the other hand, the hole-like Fermi sheet looks like flat surfaces. The multi-band nature and flatness of the sheets in Fermi surface topology are responsible to enhance the transport properties [28,51] in BKBPO perovskite.

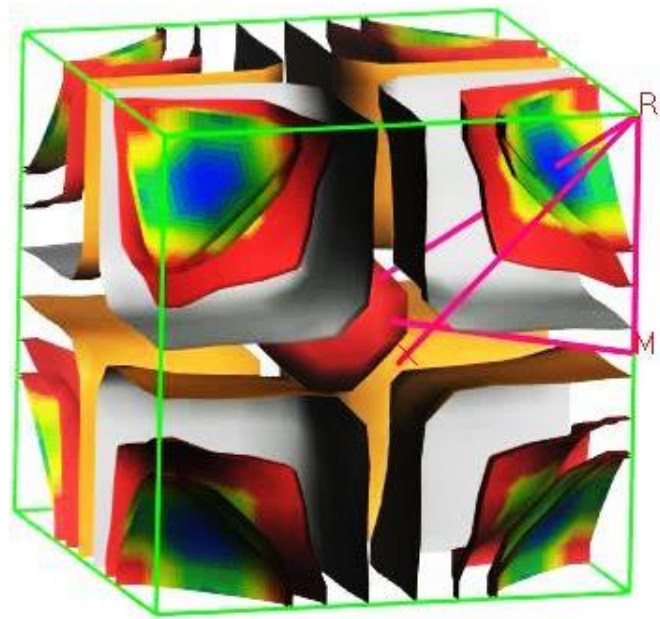

**Figure 5.** Fermi surface topology of BKBPO simple cubic perovskite superconductor.

The charge density mapping is a foremost tool to predict the bonding nature, interactions among the atoms, and exchange of electrons in space as well. The charge density mapping in the crystallographic plane of (001) with electron density scale is depicted in **Figure 6**. We also performed an electronic charge density map at different crystallographic planes with fixing $Bi^{3+}$ and $Bi^{5+}$ oxidation states and that of automatically computed valence states for comparisons. But, we didn't find any nontrivial modification under these conditions. Notably, the observed electronic charge density of BKBO in all crystallographic planes is identical, which is an indication of the isotropic nature of the material. The high and light intensity of BKBPO charge density is represented by red and blue colors, respectively. It is explicit that most charges accumulate adjacent to O and the distribution of charge is spherically uniform around O, this scenario indicates that O – Bi/Pb – O bonds are very robust. Furthermore, no charge gets transferred (overlapping) among electron distributions of Ba/K, O, and Bi/Pb ions.

We also see the strong ionic bond of O-2$p$ with Bi/Pb and K atoms from the PDOS calculation curve for the hybridization of O-2$p$ and Bi/Pb – 6$s$,6$p$ orbitals. Therefore, the Bi-based BKBPO perovskite exhibits ionic character, which is also evidenced from the mechanical property investigations. The aforementioned scenario of charge density is observed for similar perovskite superconductors. The spherical charge distribution around all atoms is connected to the ionic nature of the compound which reveals the metallic characteristics as well [25–28].

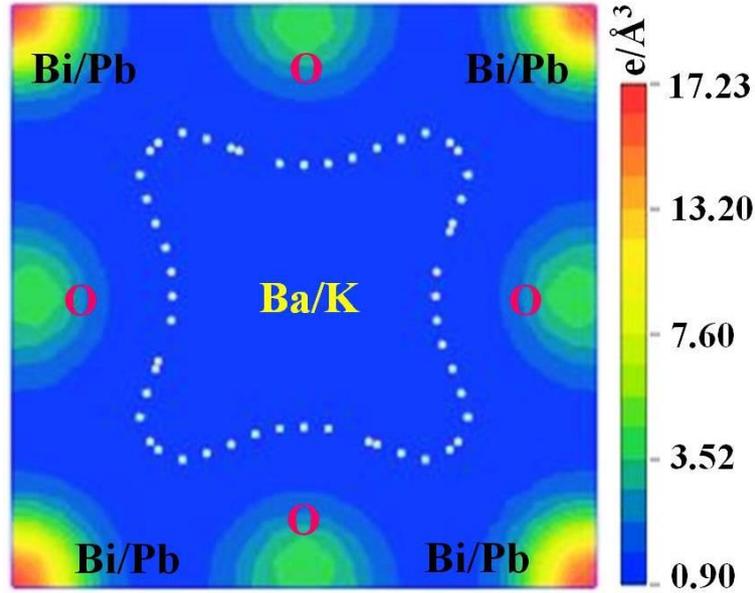

**Figure 6.** Charge density mapping of BKBPO superconductor along (001) plane.

### 3.5  *Thermodynamic properties*

The thermodynamic parameters include bulk modulus ($B$), Debye temperature ($\Theta_D$), specific heats at constant volume ($C_V$) and constant pressure ($C_P$), volume thermal expansion coefficient (VTEC), and entropy ($S$) are deeply investigated (**Figure 7**). The pressure effect is also studied in the range from 0 to 50 GPa. The bulk modulus ($B$) provides fundamental information about the bonding strength of the material, and it measures the ability to resist external deformation. The pressure dependence of the bulk modulus $B$ of the BKBPO superconductor as a function of temperature has been calculated and shown in **Figure 7(a)**. From the general physical equation of the bulk modulus ($B = \Delta P/\Delta V$), one can easily expect an increase of $B$ with pressure, because of its directly proportional relationship to the applied pressure. From **Figure 7(a)** one can see that the $B$ increases linearly with $P$, whereas, the $B$ values decrease with the increase of temperature for the particular applied pressure. It is worth mentioning that at low-temperature region ($T < 100$) bulk modulus decreases very slowly as applied pressure tends to form the structure stiffer and this nature has been clearly seen in **Figure 8(a)** for our studied superconductor. The pressure-dependent of Debye temperature, $\Theta_D$ as a function of $T$ is displayed in **Figure 7(b)**. In **Figure 7(b)**, the temperature-dependent of Debye temperature of BKBPO superconductor decreases almost linearly in the low-temperature region (T<100); further, $\Theta_D$ starts to fall. This is due to two principal effects: (i) the density of atomic oscillators declines for increasing temperature and (ii) the bonding strength in the midst of diverse atomic species also reduces. Again, $\Theta_D$ increases with the increasing pressure. This reveals that the pressure favors the improvement of the hardness of the BKBPO superconductor. We know that the Debye temperature, $\Theta_D$ is related to the maximum thermal vibration frequency of a material. Because of this relationship, the variation of $\Theta_D$ with pressure and temperature also reveals the changeable vibration frequency of the particles in BKBPO. To calculate the Debye temperature under ambient condition, we have used two different methodologies: (i) quasi-harmonic Debye approximation (QHDA) and (ii) using previously calculated elastic moduli which can be defined as **Eq. 14** [52]:

$$\Theta_D = \frac{h}{k_B}\left(\frac{3n}{4\pi V_0}\right)^{1/3} v_a \qquad (14)$$

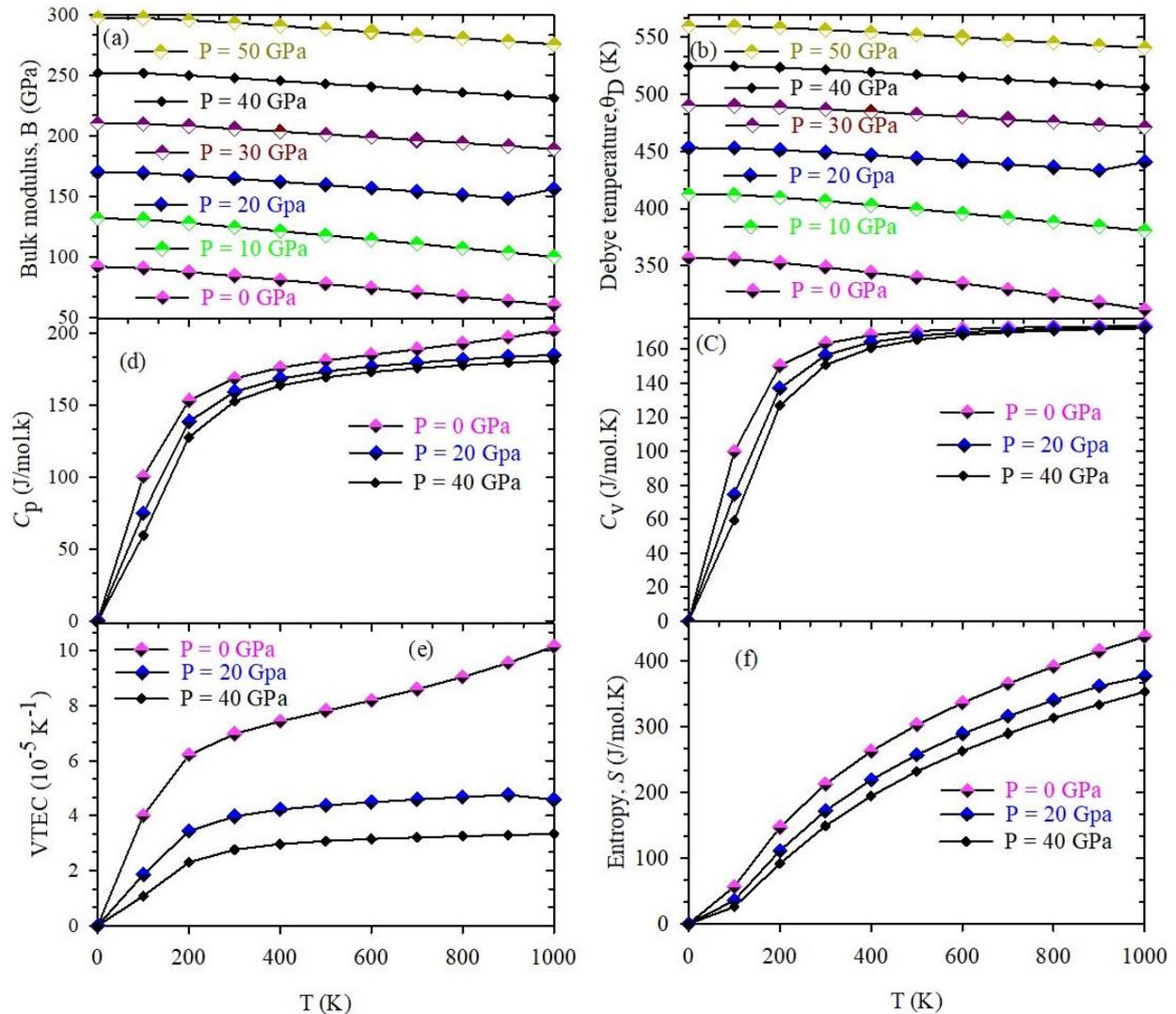

**Figure 7.** (a-f) Temperature-dependent behavior of $B$, $\Theta_D$, $C_p$, $C_v$, VTEC, and S, respectively for BKBPO superconductor under various pressures.

Our calculated values of $\Theta_D$ using the QHDA is 352.73 K, which is in good agreement with the values of 355.41 K estimated using elastic constant data (**Table 6**) giving an independent check of the accuracy [27]. Specific heat is the amount of heat required to increase the temperature of one kg of mass by 1 K. The variations of the heat capacity of the studied compound at constant volume, $C_V$, and constant pressure, $C_P$ are represented in **Figures 7(c)** and **7(d)**, respectively, with different temperatures at different pressures. From **Figure 7(c)** we observed that the specific heat at constant volume, $C_V$ increases sharply until the temperature $T = 200$ K is reached, and after this $C_V$ becomes nearly constant at high temperature. The same nature was also found for $C_P$ (**Figure 7(d)**). This indicates that phonon thermal softening originates during the increment of temperature. It can also be seen that the variations of $C_V$ and $C_P$ with temperature in the low-temperature range look similar (**Figures 7(c-d)**). But at high temperatures $C_P$ is greater than $C_V$, this can be explained by the following relationship (**Eq. 15**) [72]:

$$C_P - C_V = \alpha_V^2(T) \, BTV \qquad (15)$$

Since the variation between $C_P$ and $C_V$ is mainly determined by $\alpha_V^2$ (where $\alpha_V$ is the volume thermal expansion coefficient) for the case of low temperature. The specific heat, $C_V$ obeys the Debye law *i.e.*, $C_V \propto T^3$ in the low-temperature limit. However, at higher temperatures ($T > 300$ K), the anharmonic effect on the $C_V$ is suppressed, and it reaches a constant magnitude close to the Dulong-Petit limit (horizontal pink line), which is common to all solid products at elevated temperatures.

**Table 6.** Calculated longitudinal, transverse, average sound velocities ($v_l$, $v_t$, and $v_a$ in kms$^{-1}$), Debye temperature ($\Theta_D$ in K), melting temperature ($T_m$ in K), and electron-phonon coupling constant ($\lambda$) of BKBPO superconductor.

| Compounds | $v_l$ | $v_t$ | $v_a$ | $\Theta_D$ | | $T_m$ | $\lambda$ |
|---|---|---|---|---|---|---|---|
| | | | | using elastic moduli | using QHDA model | | |
| BKBPO | 4.298 | 2.253 | 2.517 | 355.41 | 352.73 | 1545 | 1.46 |
| BKBMO | N/A | N/A | N/A | N/A | N/A | N/A | 1.61 [27] |

The volume thermal expansion coefficient (VTEC), $\alpha_V$ as a function of temperature at different pressure is shown in **Figure 7(e)**. The expansion coefficient increases sharply at temperatures below 100 K for the BKBPO superconductor, whereas it shows a gradual slow increase at higher temperatures. Importantly, the thermal expansion coefficient is inversely connected to the bulk modulus of a solid and this nature can be seen from **Figures 7(a)** and **7(e)**. The calculated values of $\alpha_V$ at 300 K and 0 GPa for the BKBPO superconductor are $7.36 \times 10^{-5}$ K$^{-1}$. We have also calculated entropy, S for our system, and depicted the S-T diagram for different pressures in **Figure 7(f)**. Due to a decrease in volume with the applied pressure of a system, the entropy decreases with the increasing external pressure. Our studied thermodynamic parameters ($B$, $\Theta_D$, $C_p$, $C_V$, VTEC, and S) at different pressure as a function of temperature are completely comparable with the previously studied simple perovskite superconductor (Ba$_{0.62}$K$_{0.38}$)(Bi$_{0.92}$Mg$_{0.08}$)O$_3$ [27].

The melting temperature of BKBPO superconductor has also been estimated utilizing the following expression (**Eq. 16**) [64]:

$$T_m = 553 + 5.91 C_{11} \qquad (16)$$

The value of $T_m$ for the studied superconductor has been listed in **Table 6**. The larger value of melting temperature indicates the stronger bonding strength.

By using the Eliashberg equation in McMillan formula [73], we can simply calculate the electron-phonon (e-ph) coupling constant and the formula can be written as **Eq. 17**:

$$\lambda = \gamma^{expt}/\gamma^{calc} - 1 \qquad (17)$$

Where, $\gamma^{expt}$ and $\gamma^{calc}$ introduce the experimental and theoretical values of electronic specific heat coefficient, respectively. It is possible to calculate the $\gamma^{calc}$ by using the given relation [73], and then $\lambda$ can be easily found from **Eq. 17** as **Eq. 18**:

$$\gamma^{calc} = 2k_B{}^2 N(E_F)/3 \tag{18}$$

Unfortunately, no experimental data were available to address $\gamma^{expt}$ thus, we unable to estimate the magnitude of $\lambda$. Therefore, we have used the alternative McMillan's equation [74] to calculate the e-ph coupling constant (**Eq. 19**):

$$\lambda = \frac{1.04 + \mu* \ln\left(\frac{\theta_D}{1.45 T_C}\right)}{(1 - 0.62\mu*)\ln\left(\frac{\theta_D}{1.45 T_C}\right) - 1.04} \tag{19}$$

Where $\mu^*$ refers to the repulsive screened Coulomb potential. We put the calculated value of $\Theta_D = 355$ K, measured $T_c = 22.8$ K [21] and the estimated value of $\mu^* = 0.13$ (as it is empirically taken between 0.1 and 0.15 for most superconductors [75]) in the above equation, we get $\lambda = 1.46$ (**Table 6**) for BKBPO superconductor. This result indicates that the compound BKBPO is typically a strongly coupled superconductor as well. Notably, the value of the coupling constant is lower than that of isostructural BKBMO which means that the electron pairings may be reduced significantly by the substitution of Mg ions by the Pb ions in the BKBPBO simple perovskite superconductor, which lowers the magnitude of the e-ph coupling constant. The electronic band structure prediction also makes a clear difference in this connection. However, the value of $\lambda$ is similar and in between to other perovskites superconductors $(Na_{0.25}K_{0.45})(Ba_{1.00})_3(Bi_{1.00})_4O_{12}$, $(K_{1.00})(Ba_{1.00})_3(Bi_{0.89}Na_{0.11})_4O_{12}$, $(Ba_{0.62}K_{0.38})(Bi_{0.92}Mg_{0.08})O_3$, $(Ba_{0.54}K_{0.46})_4Bi_4O_{12}$ [25–28].

## 4 Conclusions

In summary, theoretically, the structural, mechanical, electronic, and thermodynamic properties of BKBPO perovskite superconductor together with its anisotropic behaviors are investigated for the first time using the CASTEP program in the framework of density functional theory. The studied single cubic perovskite entirely fulfills the Born stability criteria fairly. The elastic constants and moduli have been investigated and analyzed at zero and elevated pressures (up to 25 GPa). According to Poisson's ratio, Pugh's indicator, and Cauchy pressure, it can be concluded that BKBPO is found as brittle. Furthermore, we have also calculated the machinability index and hardness values which indicate its better industrial applications as TBC material. Metallicity in the studied compound is ensured from the band overlapping features. In the band of DOS, the O-2$p$ orbital contributes robustly at $E_F$ compared to all others. At the Fermi level, the giant dispersion for robust hybridization between orbital electrons originate flatness of the band at X point in momentum space is considered as an auspicious criterion for phonon mediated electron pairings to ameliorate superconducting behavior. We also have investigated the anisotropic nature of BKBPO perovskite, implying a greater degree of anisotropy. Besides, the presence of the electron-hole-like Fermi sheets (multi-sheets nature) and the flat faces of Fermi surfaces promote transport features in the studied superconductor. In addition, the estimated value of the electron-phonon (e-ph) coupling constant ($\lambda = 1.46$) reveals its strongly coupled superconducting nature. Finally, the quasi-harmonic

Debye approximation (QHDA) has been employed to observe the thermodynamic characteristics at high temperatures and pressures.

## Acknowledgments

Taif University Researchers Supporting Project number (TURSP-2020/45) Taif University, Taif, Saudi Arabia. This research work was also supported by grant (No. 106/5/52/R.U./Eng.) from the Faculty of Engineering, University of Rajshahi, Rajshahi 6205, Bangladesh.

## Conflict of Interests

The authors declare no conflict of interest.

## Author's contributions

**Mirza H. K. Rubel:** Conceptualization, supervision, writing manuscript and review editing; **S. K. Mitro:** Calculations, data analysis, writing manuscript draft, review editing; **Khandaker Monower Hossain:** Calculations and analysis, review editing; **Md. Mijanur Rahaman:** Conceptualization, formal analysis, review editing; **M. Khalid Hossain:** Validation and review editing; **B. K. Mondal:** Formal analysis, review editing; **Jaker Hossain:** Review editing; **Istiak Ahmed:** Calculations and analysis, manuscript drafting; **A. K. M. A. Islam:** Conceptualization and review editing; **A. El-Denglawey:** Validation, review editing, and funding.

## Data availability

All data are used to evaluate the conclusion of this study are presented in the manuscript. Additional data can be available from the corresponding authors upon reasonable request.

## References

[1] A.W. Sleight, J.L. Gillson, P.E. Bierstedt, High-temperature superconductivity in the BaPb1-xBixO3 systems, Solid State Commun. 17 (1975) 27–28. https://doi.org/10.1016/0038-1098(75)90327-0.

[2] H. Röhm, T. Leonhard, A.D. Schulz, S. Wagner, M.J. Hoffmann, A. Colsmann, Ferroelectric Properties of Perovskite Thin Films and Their Implications for Solar Energy Conversion, Adv. Mater. 31 (2019) 1806661. https://doi.org/10.1002/adma.201806661.

[3] H. Liu, H. Wu, K.P. Ong, T. Yang, P. Yang, P.K. Das, X. Chi, Y. Zhang, C. Diao, W.K.A. Wong, E.P. Chew, Y.F. Chen, C.K.I. Tan, A. Rusydi, M.B.H. Breese, D.J. Singh, L.-Q. Chen, S.J. Pennycook, K. Yao, Giant piezoelectricity in oxide thin films with nanopillar structure, Science (80-. ). 369 (2020) 292–297. https://doi.org/10.1126/science.abb3209.

[4] Y.-K. Yoo, F. Duewer, H. Yang, D. Yi, J.-W. Li, X.-D. Xiang, Room-temperature electronic phase transitions in the continuous phasediagrams of perovskite manganites, Nature. 406 (2000) 704–708. https://doi.org/10.1038/35021018.

[5] M.K. Hossain, M.C. Biswas, R.K. Chanda, M.H.K. Rubel, M.I. Khan, K. Hashizume, A review on experimental and theoretical studies of perovskite barium zirconate proton conductors, Emergent Mater. 4 (2021) 999–1027. https://doi.org/10.1007/s42247-021-00230-5.

[6] M.K. Hossain, H. Tamura, K. Hashizume, Visualization of hydrogen isotope distribution in yttrium and cobalt doped barium zirconates, J. Nucl. Mater. 538 (2020) 152207. https://doi.org/10.1016/j.jnucmat.2020.152207.

[7] M.K. Hossain, R. Chanda, A. El-Denglawey, T. Emrose, M.T. Rahman, M.C. Biswas, K. Hashizume, Recent progress in barium zirconate proton conductors for electrochemical hydrogen device applications: A review, Ceram. Int. 47 (2021) 23725–23748. https://doi.org/10.1016/j.ceramint.2021.05.167.

[8] M.K. Hossain, T. Iwasa, K. Hashizume, Hydrogen isotope dissolution and release behavior in Y-doped BaCeO 3, J. Am. Ceram. Soc. 104 (2021) 6508–6520. https://doi.org/10.1111/jace.18035.

[9] H. Wang, H. Li, S. Cao, M. Wang, J. Chen, Z. Zang, Interface Modulator of Ultrathin Magnesium Oxide for Low-Temperature-Processed Inorganic CsPbIBr2 Perovskite Solar Cells with Efficiency Over 11%, Sol. RRL. 4 (2020)

[10] J. Park, Y.-N. Wu, W.A. Saidi, B. Chorpening, Y. Duan, First-principles exploration of oxygen vacancy impact on electronic and optical properties of ABO3−δ (A = La, Sr; B = Cr, Mn) perovskites, Phys. Chem. Chem. Phys. 22 (2020) 27163–27172. https://doi.org/10.1039/D0CP05445C.

[11] C. Sun, J.A. Alonso, J. Bian, Recent Advances in Perovskite-Type Oxides for Energy Conversion and Storage Applications, Adv. Energy Mater. 11 (2021) 2000459. https://doi.org/10.1002/aenm.202000459.

[12] P. Salg, D. Walk, L. Zeinar, A. Radetinac, L. Molina-Luna, A. Zintler, R. Jakoby, H. Maune, P. Komissinskiy, L. Alff, Atomically interface engineered micrometer-thick SrMoO3 oxide electrodes for thin-film BaxSr1- x TiO3 ferroelectric varactors tunable at low voltages, APL Mater. 7 (2019) 051107. https://doi.org/10.1063/1.5094855.

[13] J.S. Manser, J.A. Christians, P. V. Kamat, Intriguing Optoelectronic Properties of Metal Halide Perovskites, Chem. Rev. 116 (2016) 12956–13008. https://doi.org/10.1021/acs.chemrev.6b00136.

[14] S.M. Kazakov, C. Chaillout, P. Bordet, J.J. Capponi, M. Nunez-Regueiro, A. Rysak, J.L. Tholence, P.G. Radaelli, S.N. Putilin, E. V. Antipov, Discovery of a second family of bismuth-oxide-based superconductors, Nature. 390 (1997) 148–150. https://doi.org/10.1038/36529.

[15] L.F. Mattheiss, E.M. Gyorgy, D.W. Johnson, Superconductivity above 20 K in the Ba-K-Bi-O system, Phys. Rev. B. 37 (1988) 3745–3746. https://doi.org/10.1103/PhysRevB.37.3745.

[16] A.W. Sleight, J.L. Gillson, P.E. Bierstedt, High-temperature superconductivity in the BaPb1−xBixO3 system, Solid State Commun. 88 (1993) 841–842. https://doi.org/10.1016/0038-1098(93)90253-J.

[17] R.J. Cava, B. Batlogg, J.J. Krajewski, R. Farrow, L.W. Rupp, A.E. White, K. Short, W.F. Peck, T. Kometani, Superconductivity near 30 K without copper: the Ba0.6K0.4BiO3 perovskite, Nature. 332 (1988) 814–816. https://doi.org/10.1038/332814a0.

[18] R.J. Cava, B. Batlogg, G.P. Espinosa, A.P. Ramirez, J.J. Krajewski, W.F. Peck, L.W. Rupp, A.S. Cooper, Superconductivity at 3.5 K in BaPb0.75Sb0.25O3: why is Tc so low?, Nature. 339 (1989) 291–293. https://doi.org/10.1038/339291a0.

[19] L.F. Mattheiss, D.R. Hamann, Electronic Structure of the High-Tc Superconductor Ba1−xKxBiO3, Phys. Rev. Lett. 60 (1988) 2681–2684. https://doi.org/10.1103/PhysRevLett.60.2681.

[20] M.H.K. Rubel, A. Miura, T. Takei, N. Kumada, M. Mozahar Ali, M. Nagao, S. Watauchi, I. Tanaka, K. Oka, M. Azuma, E. Magome, C. Moriyoshi, Y. Kuroiwa, A.K.M. Azharul Islam, Superconducting Double Perovskite Bismuth Oxide Prepared by a Low-Temperature Hydrothermal Reaction, Angew. Chemie. 126 (2014) 3673–3677. https://doi.org/10.1002/ange.201400607.

[21] M.H.K. Rubel, T. Takei, N. Kumada, M.M. Ali, A. Miura, K. Tadanaga, K. Oka, M. Azuma, M. Yashima, K. Fujii, E. Magome, C. Moriyoshi, Y. Kuroiwa, J.R. Hester, M. Avdeev, Hydrothermal Synthesis, Crystal Structure, and Superconductivity of a Double-Perovskite Bi Oxide, Chem. Mater. 28 (2016) 459–465. https://doi.org/10.1021/acs.chemmater.5b02386.

[22] M.H.K. Rubel, T. Takei, N. Kumada, M.M. Ali, A. Miura, K. Tadanaga, K. Oka, M. Azuma, E. Magome, C. Moriyoshi, Y. Kuroiwa, Hydrothermal Synthesis, Structure, and Superconductivity of Simple Cubic Perovskite (Ba 0.62 K 0.38 )(Bi 0.92 Mg 0.08 )O 3 with T c ∼ 30 K, Inorg. Chem. 56 (2017) 3174–3181. https://doi.org/10.1021/acs.inorgchem.6b01853.

[23] M. Saiduzzaman, H. Yoshida, T. Takei, S. Yanagida, N. Kumada, M. Nagao, H. Yamane, M. Azuma, M.H.K. Rubel, C. Moriyoshi, Y. Kuroiwa, Hydrothermal Synthesis and Crystal Structure of a (Ba 0.54 K 0.46 ) 4 Bi 4 O 12 Double-Perovskite Superconductor with Onset of the Transition T c ∼ 30 K, Inorg. Chem. 58 (2019) 11997–12001. https://doi.org/10.1021/acs.inorgchem.9b01768.

[24] M.H.K. Rubel, T. Takei, N. Kumada, M.M. Ali, A. Miura, K. Tadanaga, K. Oka, M. Azuma, E. Magomae, C. Moriyoshi, Y. Kuroiwa, Hydrothermal synthesis of a new Bi-based (Ba0.82K0.18)(Bi0.53Pb0.47)O3 superconductor, J. Alloys Compd. 634 (2015) 208–214. https://doi.org/10.1016/j.jallcom.2014.12.274.

[25] M.S. Ali, M. Aftabuzzaman, M. Roknuzzaman, M.A. Rayhan, F. Parvin, M.M. Ali, M.H.K. Rubel, A.K.M.A. Islam, New superconductor (Na 0.25 K 0.45 ) Ba 3 Bi 4 O 12 : A first-principles study, Phys. C Supercond. Its Appl. 506 (2014) 53–58. https://doi.org/10.1016/j.physc.2014.08.010.

[26] M.H.K. Rubel, M.A. Hadi, M.M. Rahaman, M.S. Ali, M. Aftabuzzaman, R. Parvin, A.K.M.A. Islam, N. Kumada, Density functional theory study of a new Bi-based (K1.00)(Ba1.00)3(Bi0.89Na0.11)4O12 double perovskite superconductor, Comput. Mater. Sci. 138 (2017) 160–165. https://doi.org/10.1016/j.commatsci.2017.06.030.

[27] M.H.K. Rubel, M. Mozahar Ali, M.S. Ali, R. Parvin, M.M. Rahaman, K.M. Hossain, M.I. Hossain, A.K.M.A. Islam, N. Kumada, First−principles study: Structural, mechanical, electronic and thermodynamic properties of simple−cubic−perovskite (Ba0.62K0.38)(Bi0.92Mg0.08)O3, Solid State Commun. 288 (2019) 22–27. https://doi.org/10.1016/j.ssc.2018.11.008.

[28] M.H.K. Rubel, S.K. Mitro, B.K. Mondal, M.M. Rahaman, M. Saiduzzaman, J. Hossain, A.K.M.A. Islam, N. Kumada,

Newly synthesized A-site ordered cubic-perovskite superconductor (Ba0.54K0.46)4Bi4O12: A DFT investigation, Phys. C Supercond. Its Appl. 574 (2020) 1353669. https://doi.org/10.1016/j.physc.2020.1353669.

[29] S.K. Mitro, R. Majumder, K.M. Hossain, M.Z. Hasan, M.E. Hossain, M.A. Hadi, Insights into the physical properties and anisotropic nature of ErPdBi with an appearance of low minimum thermal conductivity, Chinese Phys. B. 30 (2021) 016203. https://doi.org/10.1088/1674-1056/abaf9d.

[30] M.Z. Hasan, M. Rasheduzzaman, K. Monower Hossain, Pressure-dependent physical properties of cubic Sr B O 3 ( B = Cr, Fe) perovskites investigated by density functional theory, Chinese Phys. B. 29 (2020) 123101. https://doi.org/10.1088/1674-1056/abab7f.

[31] K.M. Hossain, S.K. Mitro, M.A. Hossain, J.K. Modak, M. Rasheduzzaman, M.Z. Hasan, Influence of antimony on the structural, electronic, mechanical, and anisotropic properties of cubic barium stannate, Mater. Today Commun. 26 (2021) 101868. https://doi.org/10.1016/j.mtcomm.2020.101868.

[32] M.H.K. Rubel, K.M. Hossain, S.K. Mitro, M.M. Rahaman, M.A. Hadi, A.K.M.A. Islam, Comprehensive first-principles calculations on physical properties of ScV2Ga4 and ZrV2Ga4 in comparison with superconducting HfV2Ga4, Mater. Today Commun. 24 (2020) 100935. https://doi.org/10.1016/j.mtcomm.2020.100935.

[33] S.K. Mitro, K.M. Hossain, R. Majumder, M.Z. Hasan, Effect of the negative chemical pressure on physical properties of doped perovskite molybdates in the framework of DFT method, J. Alloys Compd. 854 (2021) 157088. https://doi.org/10.1016/j.jallcom.2020.157088.

[34] K.M. Hossain, M. Zahid Hasan, M. Lokman Ali, Understanding the influences of Mg doping on the physical properties of SrMoO3 perovskite, Results Phys. 19 (2020) 103337. https://doi.org/10.1016/j.rinp.2020.103337.

[35] M. Rasheduzzaman, K.M. Hossain, S.K. Mitro, M.A. Hadi, J.K. Modak, M.Z. Hasan, Structural, mechanical, thermal, and optical properties of inverse-Heusler alloys Cr2CoZ (Z = Al, In): A first-principles investigation, Phys. Lett. A. 385 (2021) 126967. https://doi.org/10.1016/j.physleta.2020.126967.

[36] M.M. Rahaman, M.H.K. Rubel, M.A. Rashid, M.A. Alam, K.M. Hossain, M.I. Hossain, A.A. Khatun, M.M. Hossain, A.K.M.A. Islam, S. Kojima, N. Kumada, Mechanical, electronic, optical, and thermodynamic properties of orthorhonmbic LiCuBiO4 crystal: a first–priciples study, J. Mater. Res. Technol. 8 (2019) 3783–3794. https://doi.org/10.1016/j.jmrt.2019.06.039.

[37] K.M. Hossain, Z. Hasan, Effects of negative chemical pressure on the structural, mechanical, and electronic properties of molybdenum-doped strontium ferrite, Mater. Today Commun. 26 (2021) 101908. https://doi.org/10.1016/j.mtcomm.2020.101908.

[38] M.D. Segall, P.J.D. Lindan, M.J. Probert, C.J. Pickard, P.J. Hasnip, S.J. Clark, M.C. Payne, First-principles simulation: ideas, illustrations and the CASTEP code, J. Phys. Condens. Matter. 14 (2002) 2717–2744. https://doi.org/10.1088/0953-8984/14/11/301.

[39] M.C. Payne, M.P. Teter, D.C. Allan, T.A. Arias, J.D. Joannopoulos, Iterative minimization techniques for ab initio total-energy calculations: molecular dynamics and conjugate gradients, Rev. Mod. Phys. 64 (1992) 1045–1097. https://doi.org/10.1103/RevModPhys.64.1045.

[40] J.P. Perdew, K. Burke, M. Ernzerhof, Generalized Gradient Approximation Made Simple, Phys. Rev. Lett. 77 (1996) 3865–3868. https://doi.org/10.1103/PhysRevLett.77.3865.

[41] D. Vanderbilt, Soft self-consistent pseudopotentials in a generalized eigenvalue formalism, Phys. Rev. B. 41 (1990) 7892–7895. https://doi.org/10.1103/PhysRevB.41.7892.

[42] T.H. Fischer, J. Almlof, General methods for geometry and wave function optimization, J. Phys. Chem. 96 (1992) 9768–9774. https://doi.org/10.1021/j100203a036.

[43] H.J. Monkhorst, J.D. Pack, Special points for Brillouin-zone integrations, Phys. Rev. B. 13 (1976) 5188–5192. https://doi.org/10.1103/PhysRevB.13.5188.

[44] Z. Boekelheide, T. Saerbeck, A.P.J. Stampfl, R.A. Robinson, D.A. Stewart, F. Hellman, Antiferromagnetism in Cr3Al and relation to semiconducting behavior, Phys. Rev. B. 85 (2012) 094413. https://doi.org/10.1103/PhysRevB.85.094413.

[45] Y. Pan, M. Wen, Noble metals enhanced catalytic activity of anatase TiO2 for hydrogen evolution reaction, Int. J. Hydrogen Energy. 43 (2018) 22055–22063. https://doi.org/10.1016/j.ijhydene.2018.10.093.

[46] E. Zapata-Solvas, M.A. Hadi, D. Horlait, D.C. Parfitt, A. Thibaud, A. Chroneos, W.E. Lee, Synthesis and physical properties of (Zr 1− x ,Ti x ) 3 AlC 2 MAX phases, J. Am. Ceram. Soc. 100 (2017) 3393–3401. https://doi.org/10.1111/jace.14870.

[47] M.A. Hadi, M. Roknuzzaman, A. Chroneos, S.H. Naqib, A.K.M.A. Islam, R.V. Vovk, K. Ostrikov, Elastic and thermodynamic properties of new (Zr3−Ti )AlC2 MAX-phase solid solutions, Comput. Mater. Sci. 137 (2017) 318–326. https://doi.org/10.1016/j.commatsci.2017.06.007.

[48] C.-Z. Fan, S.-Y. Zeng, L.-X. Li, Z.-J. Zhan, R.-P. Liu, W.-K. Wang, P. Zhang, Y.-G. Yao, Potential superhard osmium

dinitride with fluorite and pyrite structure: First-principles calculations, Phys. Rev. B. 74 (2006) 125118. https://doi.org/10.1103/PhysRevB.74.125118.

[49] M.A. Blanco, E. Francisco, V. Luaña, GIBBS: isothermal-isobaric thermodynamics of solids from energy curves using a quasi-harmonic Debye model, Comput. Phys. Commun. 158 (2004) 57–72. https://doi.org/10.1016/j.comphy.2003.12.001.

[50] F. Birch, Finite strain isotherm and velocities for single-crystal and polycrystalline NaCl at high pressures and 300°K, J. Geophys. Res. 83 (1978) 1257. https://doi.org/10.1029/JB083iB03p01257.

[51] R. Majumder, S.K. Mitro, Justification of crystal stability and origin of transport properties in ternary half-Heusler ScPtBi, RSC Adv. 10 (2020) 37482–37488. https://doi.org/10.1039/D0RA06826H.

[52] R. Majumder, S.K. Mitro, B. Bairagi, Influence of metalloid antimony on the physical properties of palladium-based half-Heusler compared to the metallic bismuth: A first-principle study, J. Alloys Compd. 836 (2020) 155395. https://doi.org/10.1016/j.jallcom.2020.155395.

[53] Q. Wei, Q. Zhang, H. Yan, M. Zhang, A new superhard carbon allotrope: tetragonal C64, J. Mater. Sci. 52 (2017) 2385–2391. https://doi.org/10.1007/s10853-016-0564-6.

[54] M. Jamal, S. Jalali Asadabadi, I. Ahmad, H.A. Rahnamaye Aliabad, Elastic constants of cubic crystals, Comput. Mater. Sci. 95 (2014) 592–599. https://doi.org/10.1016/j.commatsci.2014.08.027.

[55] A. Gueddouh, B. Bentria, I.K. Lefkaier, First-principle investigations of structure, elastic and bond hardness of FexB (x=1, 2, 3) under pressure, J. Magn. Magn. Mater. 406 (2016) 192–199. https://doi.org/10.1016/j.jmmm.2016.01.013.

[56] M. Born, On the stability of crystal lattices. I, Math. Proc. Cambridge Philos. Soc. 36 (1940) 160–172. https://doi.org/10.1017/S0305004100017138.

[57] S.F. Pugh, XCII. Relations between the elastic moduli and the plastic properties of polycrystalline pure metals, London, Edinburgh, Dublin Philos. Mag. J. Sci. 45 (1954) 823–843. https://doi.org/10.1080/14786440808520496.

[58] I.N. Frantsevich, F.F. Voronov, S.A. Bokuta, I.N. Frantsevich, Elastic Constants and Elastic Moduli of Metals and Insulators, Naukova Dumka, Kiev, Ukraine, 1983.

[59] K. Shirai, A Study of Negative Force Constants: A Method to Obtain Force Constants by Electronic Structures, J. Solid State Chem. 133 (1997) 327–334. https://doi.org/10.1006/jssc.1997.7495.

[60] D.G. Pettifor, Theoretical predictions of structure and related properties of intermetallics, Mater. Sci. Technol. 8 (1992) 345–349. https://doi.org/10.1179/mst.1992.8.4.345.

[61] S.I. Ranganathan, M. Ostoja-Starzewski, Universal Elastic Anisotropy Index, Phys. Rev. Lett. 101 (2008) 055504. https://doi.org/10.1103/PhysRevLett.101.055504.

[62] M. Mattesini, R. Ahuja, B. Johansson, Cubic $Hf_3N_4$ and $Zr_3N_4$: A class of hard materials, Phys. Rev. B. 68 (2003) 184108. https://doi.org/10.1103/PhysRevB.68.184108.

[63] R. Majumder, M.M. Hossain, First-principles study of structural, electronic, elastic, thermodynamic and optical properties of topological superconductor LuPtBi, Comput. Condens. Matter. 21 (2019) e00402. https://doi.org/10.1016/j.cocom.2019.e00402.

[64] R. Majumder, M.M. Hossain, D. Shen, First-principles study of structural, electronic, elastic, thermodynamic and optical properties of LuPdBi half-Heusler compound, Mod. Phys. Lett. B. 33 (2019) 1950378. https://doi.org/10.1142/S0217984919503780.

[65] S.K. Mitro, M.A. Rahman, F. Parvin, A.K.M.A. Islam, Intermetallic $MPt_3$ (M = Ti, Zr, Hf): Elastic, electronic, optical and thermal properties, Int. J. Mod. Phys. B. 33 (2019) 1950189. https://doi.org/10.1142/S0217979219501893.

[66] D.H. Chung, W.R. Buessem, The Elastic Anisotropy of Crystals, J. Appl. Phys. 38 (1967) 2010–2012. https://doi.org/10.1063/1.1709819.

[67] X. Gao, Y. Jiang, R. Zhou, J. Feng, Stability and elastic properties of Y–C binary compounds investigated by first principles calculations, J. Alloys Compd. 587 (2014) 819–826. https://doi.org/10.1016/j.jallcom.2013.11.005.

[68] P. Ravindran, L. Fast, P.A. Korzhavyi, B. Johansson, J. Wills, O. Eriksson, Density functional theory for calculation of elastic properties of orthorhombic crystals: Application to TiSi2, J. Appl. Phys. 84 (1998) 4891–4904. https://doi.org/10.1063/1.368733.

[69] C.M. Kube, M. de Jong, Elastic constants of polycrysts with generally anisotropic crystals, J. Appl. Phys. 120 (2016) 165105. https://doi.org/10.1063/1.4965867.

[70] K. Brugger, Determination of Third-Order Elastic Coefficients in Crystals, J. Appl. Phys. 36 (1965) 768–773. https://doi.org/10.1063/1.1714216.

[71] Y.H. Duan, Y. Sun, M.J. Peng, S.G. Zhou, Anisotropic elastic properties of the Ca–Pb compounds, J. Alloys Compd. 595 (2014) 14–21. https://doi.org/10.1016/j.jallcom.2014.01.108.

[72] M.S. Ali, A.K.M.A. Islam, M.M. Hossain, F. Parvin, Phase stability, elastic, electronic, thermal and optical properties of Ti3Al1−xSixC2 (0≤x≤1): First principle study, Phys. B Condens. Matter. 407 (2012) 4221–4228. https://doi.org/10.1016/j.physb.2012.07.007.

[73] R.P. Singh, First principle study of structural, electronic and thermodynamic behavior of ternary intermetallic compound: CeMgTl, J. Magnes. Alloy. 2 (2014) 349–356. https://doi.org/10.1016/j.jma.2014.10.004.

[74] W.L. McMillan, Transition Temperature of Strong-Coupled Superconductors, Phys. Rev. 167 (1968) 331–344. https://doi.org/10.1103/PhysRev.167.331.

[75] M. Shirai, N. Suzuki, K. Motizuki, Electron-lattice interaction and superconductivity in BaPb 1-x Bi x O 3 and Ba x K 1-x BiO 3, J. Phys. Condens. Matter. 2 (1990) 3553–3566. https://doi.org/10.1088/0953-8984/2/15/012.